\documentclass[iop]{emulateapj}
\usepackage{epsfig}
\usepackage{amsmath}
\usepackage{natbib}
\usepackage{color}
\usepackage{multirow} 
\usepackage{booktabs,threeparttable}
\usepackage{psfrag}
\usepackage{graphicx}
\usepackage{subfigure}
\shorttitle{Spatial Distributions of Substructures and Dependence on Assembly History}
\shortauthors{Wang et al.}

\newcommand{\AD}[1]{#1}
\def \Msun {\ifmmode { M}_{\odot} \else ${\rm M}_{\odot}$ \fi}
\def \Mvir {\ifmmode { M_{\rm vir}} \else M$_{\rm  vir}$ \fi}

\begin{document}
\title{Satellite Alignment: I. Distribution of Substructures and their
  Dependence on Assembly History from $N$-Body Simulations}

\author{Yang Ocean Wang\altaffilmark{1,4}, W. P. Lin\altaffilmark{1,4,5}, X. Kang\altaffilmark{2}, Aaron Dutton\altaffilmark{3}, Yu Yu \altaffilmark{1,4} \& Andrea.V. Macci\`o\altaffilmark{3} }
\altaffiltext{1}{Key Laboratory for Research in Galaxies and Cosmology, Shanghai Astronomical Observatory,
Chinese Academy of Science, 80 Nandan Road, Shanghai 200030, China}
\altaffiltext{2}{The Partner Group of MPI for Astronomy, Purple Mountain Observatory, 2 West Beijing Road, Nanjing 210008, China}
\altaffiltext{3}{Max-Planck-Institute f\"ur Astronomie, K\"onigstuhl 17, 69117, Heidelberg, Germany}
\altaffiltext{4}{Graduate School, University of the Chinese Academy of Sciences, 19A, Yuquan Road, Beijing, China}
\altaffiltext{5}{Center for Astronomy and Astrophysics, Shanghai Jiaotong University, Shanghai 200240, China}
\email{wangyang,linwp@shao.ac.cn}

\begin{abstract}

  Observations have shown that the spatial distribution of satellite
  galaxies is not random, but aligned with the major axes of central
  galaxies.
  This alignment is dependent on galaxy properties, such that red
  satellites are more strongly aligned than blue satellites.
  Theoretical work conducted to interpret this phenomenon has found that it
  is due to the non-spherical nature of dark matter halos.
  However, most studies over-predict the alignment signal under the
  assumption that the central galaxy shape follows the shape of the
  host halo.
  It is also not clear whether the color dependence of alignment is
  due to an assembly bias or an evolution effect.
  In this paper we study these problems using a cosmological $N$-body
  simulation.  Subhalos are used to trace the positions of satellite
  galaxies.
  It is found that the shapes of dark matter halos are mis-aligned at
  different radii. If the central galaxy shares the same shape as the
  inner host halo, then the alignment effect is weaker and agrees with
  observational data.
  However, it predicts almost no dependence of alignment on the color
  of satellite galaxies, though the late accreted subhalos show
  stronger alignment with the outer layer of the host halo than their
  early accreted counterparts.
  We find that this is due to the limitation of pure N-body
  simulations that satellites where satellite galaxies without associated subhalos
  (`orphan galaxies') are not resolved.  These orphan (mostly red)
  satellites often reside in the inner region of host halos and should
  follow the shape of the host halo in the inner region.


\end{abstract}


\section{INTRODUCTION}

In the currently favored cold dark matter cosmology, cosmic structures
are built up of dark matter halos.  The formation of halos is
hierarchical, in that small halos form first and subsequently merge
to form bigger ones. After mergers, smaller halos become the
subhalos of the more massive host halo. Galaxies are thought to form
in the centers of halos \citep{White1978}, and most of them become
satellites when their host halos merge with a more massive
ones. After mergers, the motions of subhalos/satellites are
mainly dominated by the gravitational potential of the hosts, and in
principle they can be well traced using numerical simulations
\cite[e.g.,][]{ Springel2001} or analytical models
\cite[e.g.,][]{WhiteFrenk91, Taylor2001, Gan2010}.

It was found from $N$-body simulations \cite[e.g.,][]{Jing2002} that
dark matter halos are not spherical, but rather they are
tri-axial. The non-spherical shapes are related to the formation
history of halos, which happens preferentially along filaments.  As
the assembly history of a halo should be imprinted in the phase-space
distribution of its satellite galaxies, observational attempts have
been made to infer halo shapes using satellite distributions. Though
the task is challenging, progress has been made using the distribution
of stellar velocity \citep{Olling2000}, satellite tidal streams
\citep{Ibata2001, Lux2012, Vera2013}, and gravitational lensing
\cite[e.g.,][]{Hoekstra2004,Er2011}.

The measurements of halo shapes from satellite kinematics or weak
lensing rely on an estimate of the host potential. In fact, more
useful insight can also be gained from the pure spatial distribution
of satellites.  Most studies have focused on how satellites are distributed
with respect to the shape of the central galaxy, known as galaxy
alignment.  The observational study of the alignment of galaxies has a
long history \cite[e.g.,][]{1969ArA.....5..305H,2005ApJ...628L.101B,
  Yang2006,2007MNRAS.376L..43A, Libeskind2007, LiC2013}.  Based on large galaxy
surveys such as 2dFGRS and Sloan Digital Sky Survey, there is general agreement that the
distribution of satellite galaxies is typically along the major axis
of the central galaxy. Moreover the alignment signal depends on galaxy
color.  \citet{Yang2006} found a stronger alignment signal for red
central galaxies or red satellites. Such an effect is also seen at
high redshift \citep{Wang2010}.

A rough idea to explain the observed galaxy alignment is that if
satellite galaxies follow the distribution of dark matter, and the
central galaxy also shares a shape similar to the host halo, then the
non-spherical nature of dark matter halos naturally produces an
alignment effect. In fact, many theoretical works follow this idea
\citep{2005ApJ...629...L5, 2005ApJ...629..219Z, AB2006, Kang2005, Kang2007, Libeskind2007,
  Fal2007, Bailin2008, Fal2008, Ang2009, Fal2009, AB2010, Deason2011, 2013SSRv..177..155L}.
However, most studies over-predict the alignment signal when the
central galaxy is assumed to follow the same shape as the whole host
halo. Furthermore, most studies are unable to reproduce the alignment
dependence on galaxy color unless a dependence of central galaxy
alignment with the host halo is assumed \citep{AB2010}.

The main difficulty faced by theoretical studies of galaxy alignment
is how to assign the shape of the central galaxy.  The most natural
way is to use hydro-simulations including physics governing galaxy
formation, such as gas cooling, star formation and
feedback. Unfortunately, current simulations are typically unable to
produce a galaxy population which matches observational data (see,
however, Vogelsberger et al. 2013 and references therein).  In this
paper, we revisit the problem of galaxy alignment using an $N$-body
simulation which allows good statistics with a large number of massive
halos.  Since the simulation does not include models for galaxy
formation, we instead use the subhalos as tracers of satellite
galaxies.  For central galaxies, we follow previous studies and assume
that the shape of the central galaxy follows the shape of its host
halo. In our study, the halo shape is measured at different
iso-density surfaces using a method different from previous studies.
In an upcoming paper, we will present the results using smoothed particle hydrodynamics(SPH)
simulations performed with Gadget-2 (paper II, in preparation).

In addition to the overall alignment signal, we investigate the
dependence of alignment on the accretion and formation history of
subhalos.  Since red satellites have stronger alignment with central
galaxies \cite[e.g,][]{Yang2006}, and in general red satellites are
accreted at earlier times, it is natural to ask whether the stronger
alignment of red satellites is already set before their accretion into
the host halo or it is an evolution effect that red satellites follow
more closely the shape of dark matter halo after accretion.  To study
this question, we study the alignment of subhalos as a function of
their formation and accretion time.  To probe if the color dependence
is imprinted in the large-scale environment or is due to an evolution
effect, we also study the alignment of neighboring halos which are
within one to a few virial radii from the host halos.

The paper is organized as follows. In Section \ref{cha:axis_define},
we briefly describe the simulation and how we determine the shapes of
dark matter halos.  In Section \ref{cha:angular_distri} and
\ref{cha:mass_dependence}, we show the alignment of subhalos and its
dependence on the subhalo mass. In Section \ref{cha:time_dependence}
we investigate if the alignment signal is dependent on the accretion
or formation time of the subhalo, and present the results of alignment
of neighboring halos.  We summarize and briefly discuss our results
in Section \ref{cha:concl}.

\section{THE SIMULATION AND TRI-AXIAL HALO SHAPE} \label{cha:axis_define}

The cosmological simulation used in this paper was performed using the
massive parallel code Gadget-2 \citep{Springel2001,
  2005MNRAS.364.1105S} and evolved from redshift $z=120$ to the
present epoch in a cubic box of $100 {\rm Mpc} h^{-1}$ with $512^3$ dark
matter particles, assuming a flat $\Lambda{\rm CDM}$ ``concordance''
cosmology with $\Omega_{\rm m}=0.268$, $\Omega_{\Lambda}=0.732$,
$\sigma_8=0.85$. A Plummer softening length of $4.5 {\rm kpc}$ was
adopted and each dark matter particle has a mass of about
$5.5\times10^8 h^{-1}{ M_{\odot}}$. The simulation has been widely
used in previous studies \cite[e.g.,][]{Jing2006, Zhu2006, Lin2006}
and the readers are referred to these papers for more details of the
simulation.  In total 62 snapshots between $z=15$ and $z=0$ have been
used to construct the merger history of halos and subhalos. All halos
were found using the standard friends-of-friends algorithm with a
linking length of 0.2 times the mean particle separation, while
subhalos were found using the SUBFIND routine
\citep{2001MNRAS.328.726S}.

\AD{ We select host halos as those with a mass larger than
  $10^{12}M_{\odot}$ (to mimic the data of Yang et al. 2006). In our
  simulation there are about 2000 host halos in total, at redshift
  $z=0$.}  \AD{Then we re-select host halos with $\theta_{\rm err}$ of
  inner axes less than $5^{\circ}$. Here $\theta_{\rm err}$ is the
  error on determining the axis given by following equation
  \begin{equation}
    \theta_{\rm err}=\frac{1}{2\sqrt{N}}\frac{\sqrt{r}}{1-r},
  \end{equation}
  where $N$ is the number of particles used and $r$ is the relevant axis
  ratio: $b/a$ for the major axis, $c/b$ for the minor axis, and
  max$(b/a, c/b)$ for the intermediate axis (here $a$,$b$,$c$ are the
  lengths of the axes with $a\ge b \ge
  c$)\cite[cf.][]{2004ApJ...616...27B}.  There are 1604 halos that
  survive our two cuts. In Table.~\ref{tab:1} we list the number of
  host halos in different mass bins. Unless otherwise specified, the
  following analysis will use the same halo catalogue as described in
  Table.~\ref{tab:1}.}

There are various methods to determine the shape of halos and they
differ in details \citep{Bailin2005}. In the most widely used method,
the axes of a dark matter halo are calculated by its overall inertia
tensor defined as
\begin{equation}
  I_{ij}=\sum\nolimits_nx_{i,n}x_{j,n} ,
\end{equation}
where $x_{i,n}$ is the distance of particle $n$ from the halo center
in dimension $i$ and the summation is over all particles in a halo.
In the three-dimensional(3D) configuration space $i,j=1,2,3$.  The three eigenvectors of
$I_{ij}$ define the orientation of the three axes of a halo, and the
\AD{eigenvalues are related to the rms along the corresponding
  eigenvectors}.  The halo shape determined in this way should be
closely correlated with the large scale environment of the halo, for
example, external tidal field or large nearby filaments
\AD{\citep{2002MNRAS.332..339P}}.

\AD{An improved version of the above method uses the reduced inertia
  tensor defined as
  \begin{equation}
    I_{ij}=\sum\nolimits_n\frac{x_{i,n}x_{j,n}}{{r_n}^2},
  \end{equation}
  which can alleviate the contamination of substructures
  \citep{1983MNRAS.202.1159G, Bailin2005}.  Here ${r_n}$ is the
  distance of the $n$th particle from the halo center.  In this work,
  when we calculate the axes for the whole halo, we use this reduced
  inertia tensor.}

Since dark matter halos form hierarchically, the inertia tensors of
the whole halo might not precisely describe its internal orientation
or that of the central galaxy residing in it.  Therefore, following
\citet{Jing2002} we use different iso-density surfaces to determine
the axes of a halo. The local density at a particle position is
calculated by using the smoothed-particle-hydrodynamics (SPH) method.
We select those particles with local density within a narrow bin as an
input list for the group finder and apply the FoF method again to find
subhalos using a smaller linking length.  To exclude particles from
subhalos and eliminate their influence on the shape determination of
the host halo, only those particles within the largest sub-group are
retained.  Finally the iso-density surfaces of a halo are built up and
then the axes of them can be calculated by fitting with a tri-axial
ellipsoid.

\AD{Following \citet{Jing2002}, we determine the axes for five
  different iso-density surfaces, for which the $n$th bin has local
  density of $100 \times 5^{n-1} $ ($n=1,2,3,4,5$) times the cosmic
  critical density.  The mean radius of each elliptical surface is
  approximately 0.6, 0.4, 0.25, 0.12, and 0.06 of the virial radius
  respectively.  The orientation of the five iso-density surfaces
  represents the orientation of different parts of a halo, from the
  most inner part to the outer part.}  For convenience, in the
following content we will call the axis of the most outer part of the
halo as ``\textit{outer axis}''($n$=1), the axis of the most inner part
of the halo as ``\textit{inner axis}''($n$=5) and the axis of
intermediate part as ``\textit{intermediate axis}''($n$=3).

\begin{figure}
\centerline{\psfig{figure=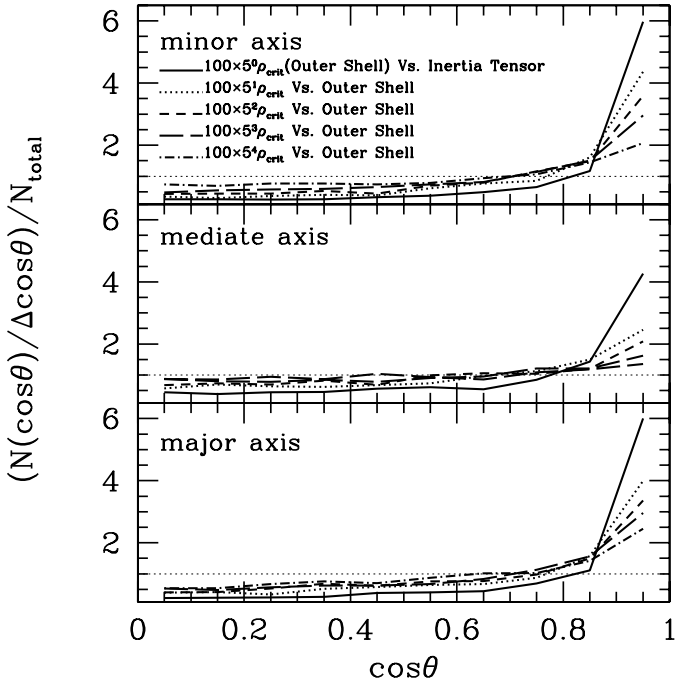,width=0.5\textwidth}}
\caption{Comparison of the angle between the axes determined from two
  different methods (iso-density and reduced inertia tensor). The top,
  middle and bottom panels are for the minor, mediate, and major axes,
  respectively. The solid lines show the angle between the axes of the
  outer shell (from the iso-density method) and that determined from
  the inertia tensor. Other lines are the alignment between the
  different shells using the iso-density method. }
\label{axis_diverse}
\end{figure}

Intuitively, the shape of the outer iso-density ellipsoids should be
closely correlated to that of the whole halo. In
Figure~\ref{axis_diverse} we plot the alignment between the axes of the
iso-density ellipsoid and that of the halo inertia tensor.  It is not
surprising to find that the outer axes align strongly with the axes
derived from the inertia tensor of the whole halo, and that the
strength of alignment for the inner axes becomes weaker (but still
significant).  Interestingly but not unusual, it again commendably
demonstrates the hierarchical structure of dark matter
halos. The results also show that there is non-negligible
mis-alignment between the halo major `\textit{inner axis}' and major
`\textit{outer axis}', which may imply a mild alignment signal
between BCG orientation and satellite distribution, as presented in
the next section. The alignment between the `inner' and `outer' axes
is stronger with increasing halo mass, with average alignment angles
of $46^{\circ}.2, 38^{\circ}.8, 26^{\circ}.0$, for host halos with $
10^{12} \Msun < \Mvir < 10^{13} \Msun$, $ 10^{13} \Msun < \Mvir <
10^{14} \Msun$, and $\Mvir \geq 10^{14} \Msun$, respectively.
\AD{Although determining axes will suffer from the limited number of
  particles especially for small halos, our strategy which excludes
  host halos with less accuracy ($\theta_{\rm err}>5^{\circ}$) can
  alleviate this problem and making the above results more reliable.}

\section{SPATIAL DISTRIBUTION OF SUBHALOS AND NEIGHBORING HALOS} \label{cha:angular_distri}

In this section, we present the spatial distribution of subhalos and
the neighboring halos from our simulation. For each host halo, we
place a central galaxy at its center whose shape is set to follow the
shape of the host dark matter halo (either inner, intermediate or
outer axis). The neighboring halos are defined as individual halos
which are less massive than the host halo and reside within the range
of $1-3r_{\rm vir}$ from the center of the host halo.  To obtain more
robust results, we limit subhalos to being more massive than
$1.65\times10^{10}h^{-1}M_{\odot}$ (containing more than 30 particles)
and neighboring halos to being more massive than
$5.5\times10^{10}h^{-1}M_{\odot}$ (containing more than 100
particles). \AD{Table ~\ref{tab:1} lists the number of subhalos and
  neighboring halos in each mass bin.}

\begin{table}[htbp]
  \centering
  \begin{threeparttable}
    \caption{Number of host halos in different mass bins, and subhalos
      and neighboring halos with different masses of their host or of
      themselves. }
    \begin{tabular}{cccc}
    \toprule
    Range of $M_{\rm host}/[h^{-1}$M$_{\odot}]$ \tnote{a}& $10^{12} \sim 10^{13}$ & $10^{13} \sim 10^{14}$ & $\ge 10^{14}$ \\
    \hline
    Host Halo & 1206 & 374 & 24\\
    Subhalo & 4193 & 7175 & 3793\\
    Neighboring Halo & 1703 & 3238 & 1625\\
    \hline
    Range of $M_{\rm self}/[h^{-1}$M$_{\odot}]$ & $10^{10}\sim10^{11}$ \tnote{b} & $10^{11}\sim10^{12}$ & $\ge 10^{12}$ \\
    \hline
    Subhalo & 11609& 3178 & 374\\
    Neighboring Halo & 2961 & 3249 & 356\\
    \bottomrule
    \end{tabular}
    \label{tab:1}
    \begin{tablenotes}
    \item[a] Host halos with inner axes less accurate than
      $5^{\circ}$ and their subhalos and neighboring halos are
      excluded from this table.
    \item[b] Note that we set a minimum number of 30 and 100 particles
      for subhalos and neighboring halos, so
      $10^{10}-10^{11}h^{-1}M_{\odot}$ means about
      $1.65\times10^{10}-10^{11}h^{-1}M_{\odot}$ for subhalos and
      $5.5\times10^{10}-10^{11}h^{-1}M_{\odot}$ for neighboring
      halos.
    \end{tablenotes}
  \end{threeparttable}
\end{table}

\begin{figure*}
  \centerline{\psfig{figure=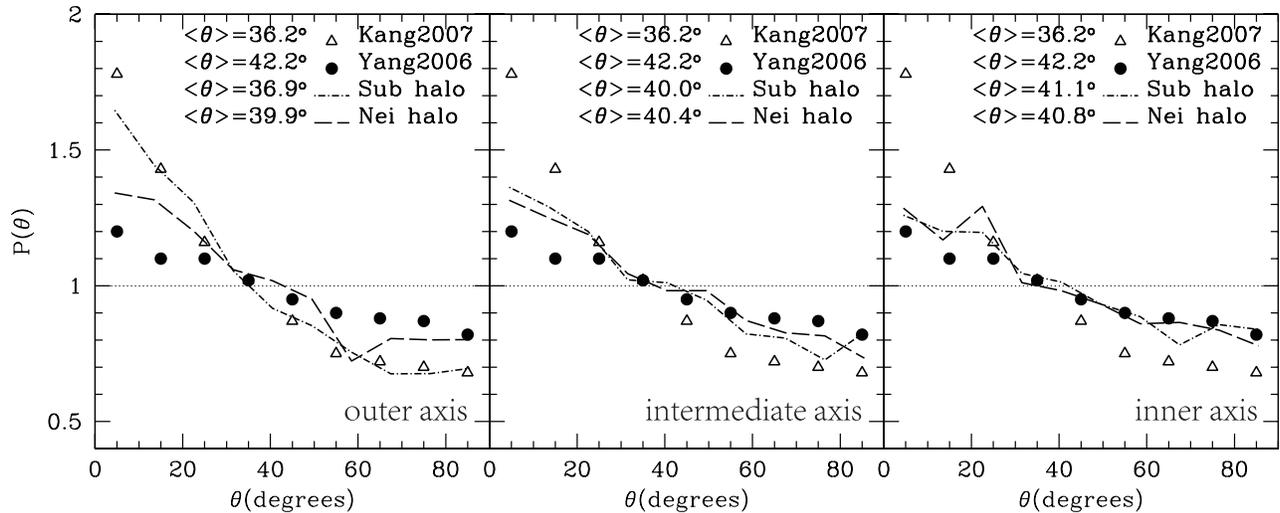,width=0.95\textwidth}}
  \caption{Angular distribution of sub-halos and neighboring
  halos along the major axis of host halo, converted to
  a format compatible with Yang and Kang's work. The short and long dashed lines
 represent the sub-halos and neighboring halos respectively.
 The open triangles are theoretical results of \citet{Kang2007} while
 the solid dots are the observations of \citet{Yang2006}. From the
 left to the right panels, the lines represent statistical results
 using the major axis of the outermost part, intermediate part and
 innermost part of halos, respectively. The horizontal dotted line indicates
 a random distribution. For convenience, the legend ``neighboring'' is
 replaced with ``Nei'' here and in all plots throughout.}
\label{ang_comp}
\end{figure*}

The alignment signal is described as the distribution of angular
separation between the axis of the host halo and the connecting line
between the centers of the host halo and each subhalo/neighboring
halo. The distribution is described as,
\begin{equation}
  P(\theta)=\frac{N(\theta)}{\langle N_R(\theta)\rangle},
\end{equation}
where $N(\theta)$ is the count of subhalos (or neighboring halos) in
bin $\theta$, while $\langle N_R(\theta)\rangle$ is the average count
of random samples in the same bin. So $P(\theta)=1$ means the absence
of any alignment, while $P(\theta)>1$ at small $\theta$ implies a
distribution with a preferred alignment along the major axis.

\AD{To compare with the observational results of Yang et al. (2006),
  here we performed the analysis in two-dimensional space. We first calculate the axes in
  3D space, and then project them onto the $X-Y$ plane. The angular
  position is then obtained for every subhalo/neighboring halo using
  its projected position in the same plane. To generate the random
  sample, for each host halo containing $N$ subhalos), we produce $N$
  subhalos in 3D space with a spherical random distribution. In
  principle, the produced random sample should share the same radial
  distribution of the real subhalos in each host halo. However, as our
  calculation is only dependent on the angular separation, our results
  are not affected by this requirement. }

In Figure~\ref{ang_comp} we show the alignment along the major axes of
central galaxies. The left, middle and right panels show the results
assuming that the central galaxies follow the shape of their host
halos defined at the outer, intermediate and inner axes.  In each
panel the dotted and dashed lines are for subhalos and neighboring
halos, respectively. For comparison with other results, we also plot
the observational work of \cite{Yang2006}, and the theoretical one by
\cite{Kang2007} as circles and triangles in each panel.

Figure~\ref{ang_comp} shows that both subhalos and neighboring halos are
aligned with the major axes of central galaxies.  The signal is
stronger than the data (solid points) if the central galaxy follows
the shape of the dark matter halo at the outer or intermediate axes
(left and middle panels). Our results are consistent with those of
Kang et al. (2007), who used satellite galaxies from their Semi-analytic models(SAMs) and
have found strong alignment if central galaxy follows the shape of the
whole dark matter halo.  \cite {Kang2007} pointed out that some
observational effects, such as the flux-limit, redshift-space
distortions and the galaxy group finder, could cause a shallowing of
the alignment signal, but these would not be enough to reconcile the
simulations with observations.  They further proposed that the
observed alignment signal could be reproduced if the spin of the
central galaxy aligns with that of dark matter halo.

Our result shows that there may be another way to remove such a
discrepancy with observations. As showed in the right panel of
Figure~\ref{ang_comp}, the predicted alignment signal is close to the
data if the shape of the central galaxy follows the shape defined at
the inner halo region.  This is also consistent with the results of
\cite{Fal2009} in which they also found only a weak misalignment if
using the inner halo to define the shape of central galaxy, although
they use a different method to measure the halo shape.

It is also found from Figure~\ref{ang_comp} that the alignment of
neighboring halos is similar to that of the subhalos.  There seems to be a
slight evolution effect that subhalos' alignment is stronger if the
central galaxy follows the outer shape of dark matter halo (left
panel). However, the predicted alignment signal is too strong in this
case, thus it can not account for the observed color dependence of
satellite alignment. If the central galaxy follows the shape of the
inner axis, and blue satellite galaxies are recently accreted from
neighboring halos, then their alignment should be identical to that of
the red satellites (subhalos in our case).  To explain the observed
color dependence, it will require that either blue (red) satellites
are not random sample of neighboring halos (subhalos). We will later
investigate if the color dependence arises from the assembly bias of
satellite galaxies.

The right panel of Figure~\ref{ang_comp} shows that if the central
galaxy follows the inner shape of the dark matter halo, then the
predicted alignment signal is more or less consistent with the
data. However, it is not implied that this is the only way to
reproduce the observed alignment signal.  It indicates that some
degree of misalignment between the central galaxy and the overall
shape of dark matter halo has to be assumed.  For example,
\cite{Kang2007} and \cite{AB2010} both have found that if the minor
axis of central galaxies follows the angular momentum of the dark
matter halo, then the alignment of satellites is also closer to the
observational data.

\begin{figure}
  \centerline{\psfig{figure=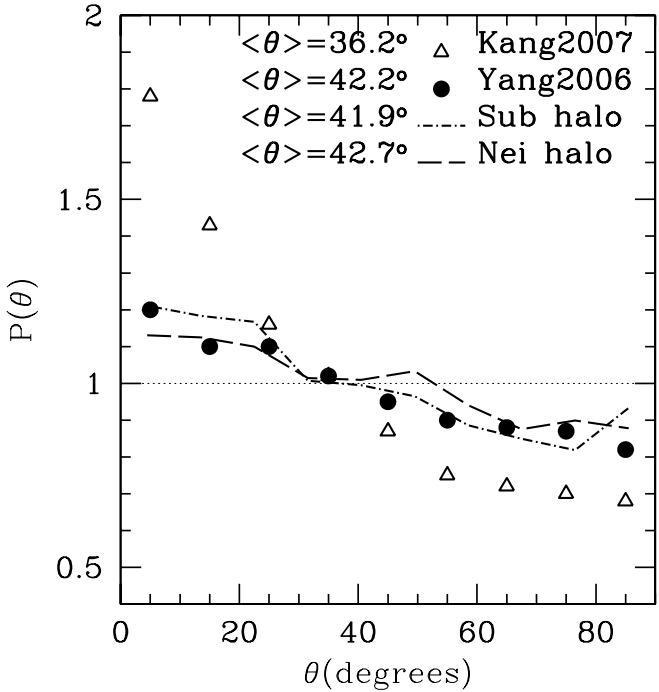,width=0.5\textwidth}}
  \caption{The angular distribution of sub-halos and neighboring halos
    along a designed axis of center galaxy.  See the main text for
    more details.  }
  \label{ang_comp_re}
\end{figure}

The observed alignment effect can also be reproduced if the central
galaxy follows the overall shape of the dark matter halo, but with
some distribution. Figure~\ref{ang_comp_re} shows the results when the
angle between the major axes of the central galaxy and the host halo
follows a Gaussian distribution with a mean of $0^{\circ}$ and a
standard deviation of $25^{\circ}$. It is seen that the observed
alignment signal is well reproduced. \AD{Some previous works find
  results consistent with this: \citet{Bailin2005} found a
  misalignment between the angular momentum vector and the minor axis,
  with a mean value of about $25^{\circ}$; \citet{2010MNRAS.404.1137B}
  found that there is a median angle of $25^{\circ}$ between the
  angular momentum vectors of inner ($\le 0.25R_{\rm vir}$) part and
  the whole halo.}  In fact, there is observational evidence for a
mis-alignment angle of about $23^\circ$ between the major axis of
brightest central galaxies (BCG) and that of their dark matter halos
inferred from the distribution of satellite
galaxies\citep{2008MNRAS.385.1511W}.  In addition, a larger
mis-alignment angle of about $35^\circ$ was also suggested from
gravitational shear-intrinsic ellipticity correlation of luminous red
galaxies \citep{2009ApJ...694L..83O, 2009ApJ...694..214O}.

So far, there is no definitive way to assign a shape to the central
galaxy. The most promising way should be the one which can reproduce
not only the overall alignment of satellites, but their dependence on
galaxy properties.  Studies based on $N$-body simulations and
semi-analytical models for galaxy formation \AD{\citep{Kang2007,
    AB2010, Libeskind2005, 2005ApJ...624..505Z}} have made some
progress to achieve this goal.  However, these studies still face
problems.  First, the predicted total alignment signal is still
stronger than observations.  Second, the color dependence of
satellites in the models themselves are not correctly reproduced
\cite[however, see][]{AB2010}. Third, it is difficult to interpret the
contamination of interlopers in observations.

In our simulations we have no galaxies, only subhalos. Thus we are
unable to explore the dependence of alignment on galaxy
properties. However, as galaxy properties are closely related to the
halo (subhalo) mass and accretion time \citep{Yang2012}, we could
learn the shape of the central galaxy from the dependence of alignment
on the mass of host halos and subhalos. In particular, we want to
study if the alignment is dependent on the accretion/formation time of
subhalos and will present such results in the following sections.

\section{DEPENDENCE ON HALO/SUBHALO MASS}\label{cha:mass_dependence}

\begin{figure*}
  \centerline{\psfig{figure=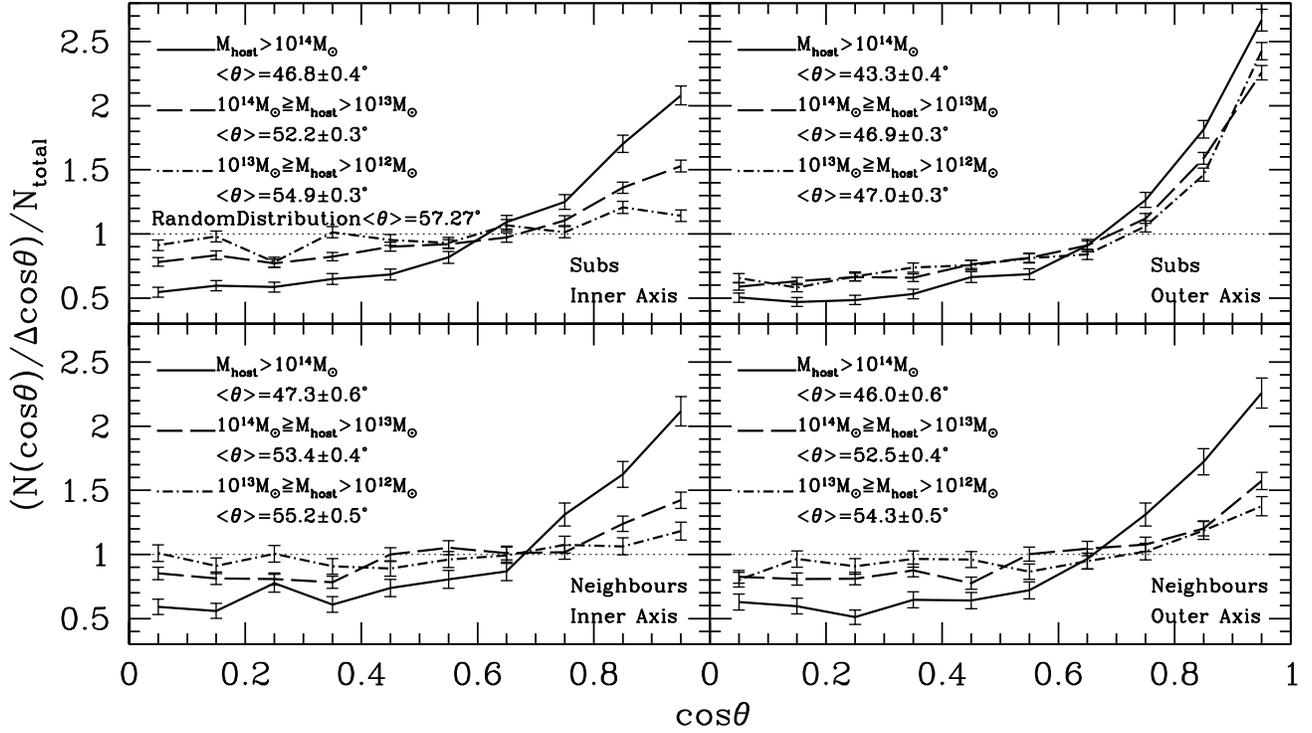,width=0.95\textwidth}}
  \caption{The dependence of the angular distribution for sub-halos
    (upper panels) or neighboring halos (bottom panels) on host halo
    mass. The vertical coordinate represents the probability while the
    horizontal coordinate represents the cosine of the angular
    separation (see main text for definition). The alignment along
    ``\textit{inner axis}'' and ``\textit{outer axis}'' are plotted in
    the left and right panel respectively. \AD{Error bars indicate 1
      $\sigma$ standard deviation in each bin.}  Different lines
    represent different bins of host halo mass. An isotropic angular
    distribution corresponds to the horizontal dotted lines.}
  \label{ang_host}
\end{figure*}

Observationally, it was found that the alignment of satellites is
stronger in massive halos, and it is even stronger for red satellites
in massive red central galaxies \cite[c.f.,][]{Yang2006}. Such a mass
dependence should be imprinted in the alignment of
subhalos/neighboring halos.  In this section, we investigate the
dependence of alignment on the host halo mass and the mass of
subhalos/neighboring halos themselves.  Different from the previous
section, here we give the results in 3D space since it gives a
stronger alignment signal in real space.  Figure~\ref{ang_host} shows
the effects of host halo mass on the alignment of subhalos and
neighboring halos.  Note that the error-bars are for the $1 \sigma$
standard deviation of the mean throughout the paper. It is found that
more massive host halos have a stronger alignment signal.  Though
subhalos/neighboring halos preferentially align along the
``\textit{outer axis}'', it seems that dependence on halo mass is
stronger for the alignment along the ``\textit{inner axis}''.  This
reflects the dependence of central-to-outer alignment strength with
halo mass.  Given the fact that more massive BCGs have redder colors,
the results for ``\textit{inner axis}'' are in agreement with
observational results of stronger alignment for red central galaxies
than blue central galaxies \citep{Yang2006}, as well as the results
from a semi-analytical model \cite[e.g.,][]{Kang2007}.

\begin{figure*}
  \centerline{\psfig{figure=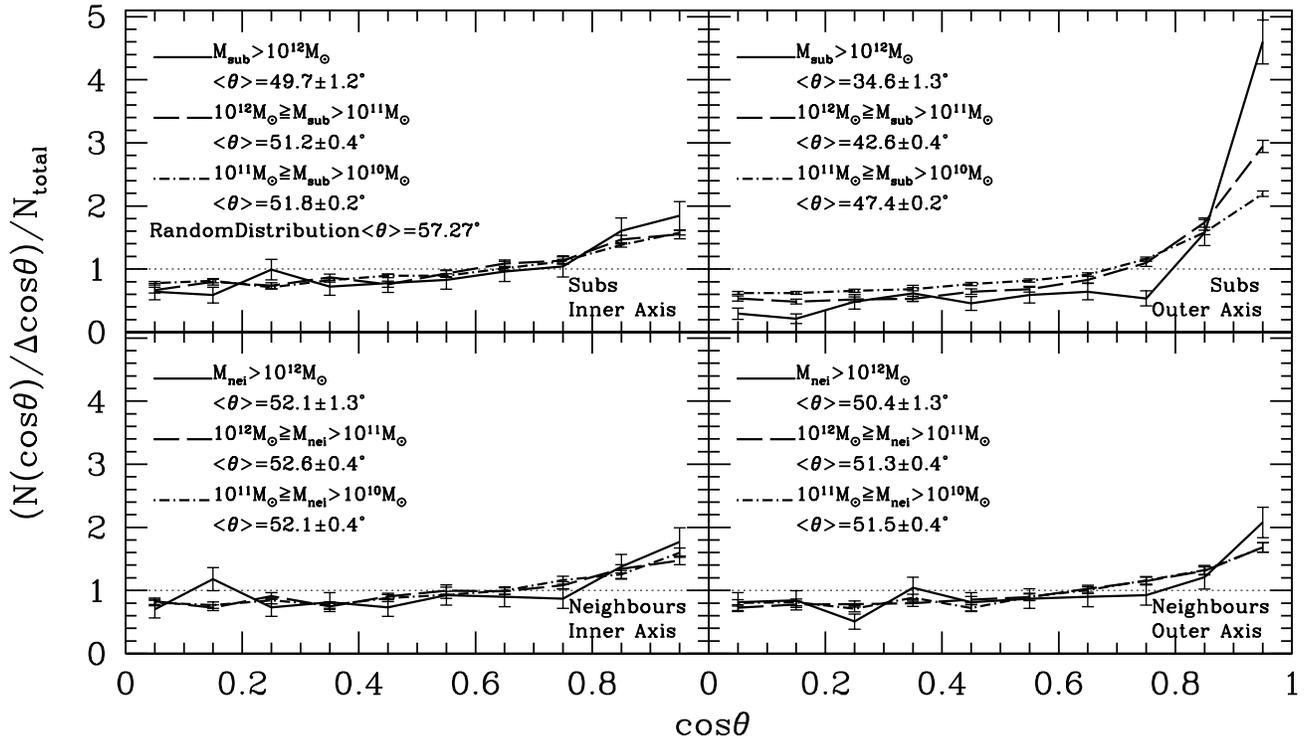,width=0.95\textwidth}}
  \caption{The dependence of angular distribution on the mass of
    sub-halos or neighboring halos. Other description is the same as
    Figure~\ref{ang_host} except that different lines are for mass bins
    of sub-halos/neighboring halos.}
  \label{ang_self}
\end{figure*}

In Figure~\ref{ang_self} we show the alignment of subhalos/neighboring
halos with dependence on their own mass. Contrary to the dependence on
the host halo mass, the dependence is quite weak on the mass of
subhalos/neighboring halos themselves.  Only those most massive
subhalos/neighboring halos show a little stronger alignment signal
along the ``\textit{outer axis}'' of their host halo than less massive
ones. We also find that if the mass of subhalos at accretion is used
instead, the results are also very similar.  Assuming that massive
subhalos or neighboring halos harbor redder satellites and the major
axis of the central galaxy can be represented by the ``inner axis'' of
host halos, the results of alignment along the ``inner axis'' (left
panels in Figure~\ref{ang_self}) can not explain the observational trend
that red satellites are better aligned than the blue counterparts
\citep{Yang2006}.

Our results above clearly show that the neighboring halos are
preferentially distributed along the major outer axes of the host
halos. This is easy to understand as the neighboring halos define the
local large-scale environment of the host halo which dominate the
direction of mass accretion. Also found is that more massive host
halos are stronger aligned with the neighboring halos (lower right
panel of Figure~4) and the most massive neighboring halos have stronger
alignment with the host halos (lower right panel of Figure~5).  These
results are consistent with the picture that more massive host halos
are connected with stronger filaments where more massive subhalos are
embedded and accreted along them.

\section{DEPENDENCE ON ASSEMBLY HISTORY}\label{cha:time_dependence}

Intuitively one would speculate that galaxies which formed earlier
(such that their host halos formed earlier) should have a redder
color, therefore there could be helpful clues for the dependence of
alignment signal on galaxy color from the formation histories of
halos.  On the other hand, being accreted is an important event for a
subhalo. One would expect that subhalos accreted earlier will spend
more time within host halos and suffer longer from tidal stripping as
well as dynamical friction and thus may lose memory of initial
accretion positions(as \cite{2005MNRAS.364..424W, 2007MNRAS.375..633W}
indicate, these initial accretion positions are more likely to be close
to the major axis of the host halo), consequently attenuating the alignment
signal. However, it is also possible that earlier accreted satellites
will experience more gravitational effects from the host, thus
following the shape of the host more closely and predicting stronger
alignment \citep{Yang2006}.  These two effects will compete with each
other, and finally determine the alignment of satellite galaxies.

In this section, we focus on the relation between subhalos' (or
neighboring halos') angular alignment and their assembly history.
Here we define two key time points for the formation and evolution of
subhalos.  A subhalo was once an independent halo before it is
accreted into a virialized halo nowadays. We define the time when the
progenitor of a subhalo was accreted into a large halo as ``{\it
  accretion time}''.  \AD{As subhalos often cross the virial radius of
  the host halos for a few times \cite[e.g.][]{2009ApJ...692..931L},
  we use the first time it cross the virial radius as the accretion
  time.} Before being accreted, the progenitor halo evolves as an
individual halo, and its mass grows almost monotonically with time
\citep{DeL2004}.  The time when this progenitor halo has assembled
half of its mass at the accretion time is defined as its ``{\it
  formation time}''. For neighboring halos, there is no accretion time
to apply since it has not been accreted by the central
halo. \AD{Therefore we define their ``{\it formation time}'' as the
  time when they accrete half of their mass at $z=0$ and consider the
  effect of their ``{\it formation time}'' only.}

\begin{figure}
  \centerline{\psfig{figure=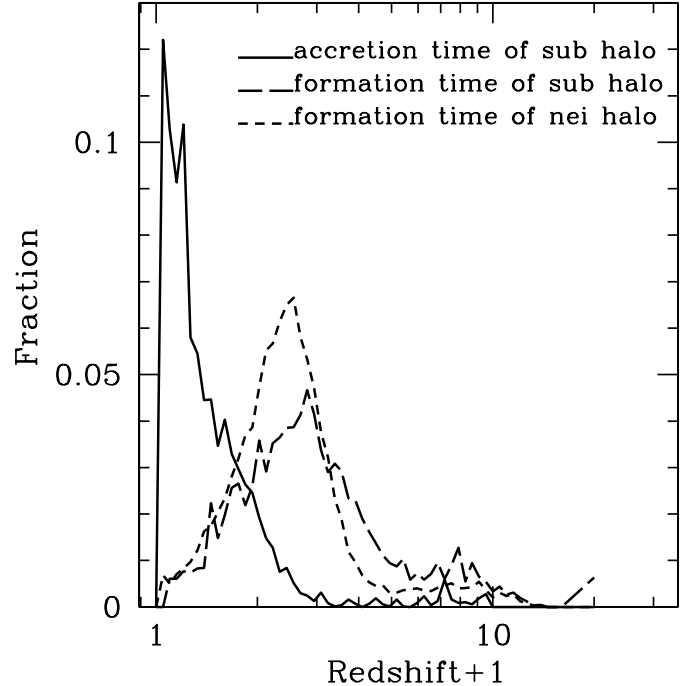,width=0.5\textwidth}}
  \caption{The distribution of formation and accretion
    redshifts. Solid line represents the distribution of sub-halos'
    accretion time, while long and short dashed lines are for the
    formation time of sub-halos and neighboring halos
    respectively. \label{time_distri}}
\end{figure}

First, we show the distribution of accretion and formation redshifts
of \AD{all subhalos/neighboring halos in our sample} in
Figure~\ref{time_distri}. The accretion redshifts mostly range from $z=0$
to $z=5$. One can find that most subhalos are accreted to their host
halos at redshift $z<1$.  This is consistent with previous work
\citet{DeL2004}. Note that here we only analyze the accretion
redshifts for those subhalos which finally survive at $z=0$ in their
host halos.  In some previous works \cite[e.g.,][]{Yang2011}, the
statistic of accretion time based on merger trees including
un-resolved subhalos are different, which have more accretion events
at higher redshift. Also in our simulation there are some subhalos
which can not be traced back to their individual halo progenitors due
to the resolution of the simulation. Such kind of subhalo accounts for
approximately $15\%$ of all subhalos.  The formation time distribution
of neighboring halos is consistent with the prediction for low mass
halos \cite[e.g.,][]{Lin2003}.

\begin{figure}
\centerline{\psfig{figure=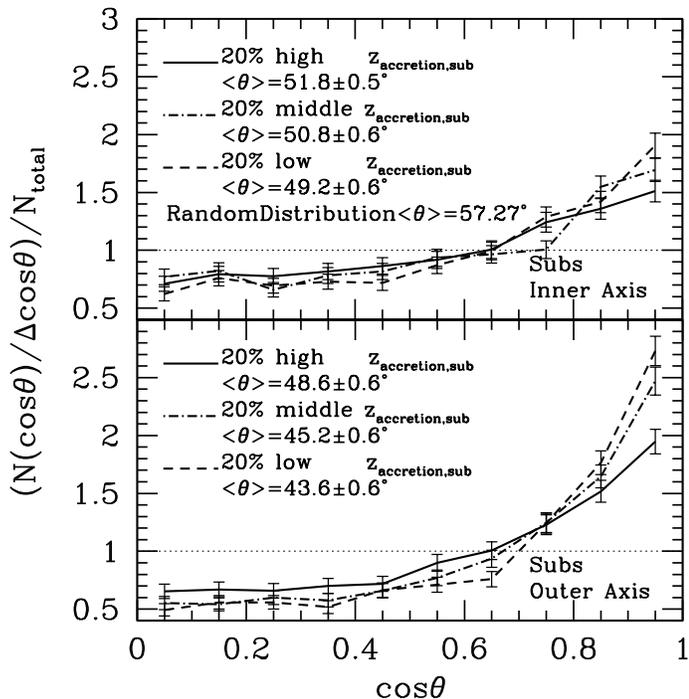,width=0.5\textwidth}}
\caption{The dependence of angular distribution on accretion
  redshift. The solid line, dotted dashed line and dashed line are for
  sub-halos with relatively high, middle and low redshift accretion
  time respectively. The horizontal dotted line represents an
  isotropic angular distribution. \AD{Error bar indicates 1 $\sigma$
    standard deviation in each bin.} In each panel, each redshift bin
  contains 20\% of all samples.}
\label{accretion}
\end{figure}

\AD{Figure~\ref{accretion} shows the angular distribution of subhalos
  with different accreted redshifts. We choose the 20\% of subhalos
  accreted earliest, the 20\% accreted latest and 20\% accreted in
  mid-period from the whole samples and plot their angular
  distribution along the major axes of their host halos.}  Note that
the use of 20\% is arbitrary.  We have also tried using choosing
criterion of 25\%, 33\% and 40\% (not shown in the plots). As might be
expected, it turns out that the difference between early accreted
sub-halos and late accreted ones becomes less distinct when the
samples include more intermediate accretion redshifts.

The upper panel of Figure~\ref{accretion} shows that the alignment
signals along the inner axes of host halos are almost identical
(within the 1$\sigma$ errors) for the three samples with different
formation times. The lower panel shows that later accreted subhalos
have significantly stronger alignment with the outer axes of the host
halos. From models of galaxy formation \cite[e.g.,][]{Kang2008,
  Guo2011}, it is expected that early accreted subhalos should host
red satellite galaxies.  The observed stronger alignment of red
satellites implies that early accreted subhalos should have strong
alignment. Indeed such a dependence is seen from the analysis of
\cite{AB2010}.  However, such an expectation is not seen, and is in fact contrary
to what we see from Figure~\ref{accretion} in our work.

\begin{figure}
  \centerline{\psfig{figure=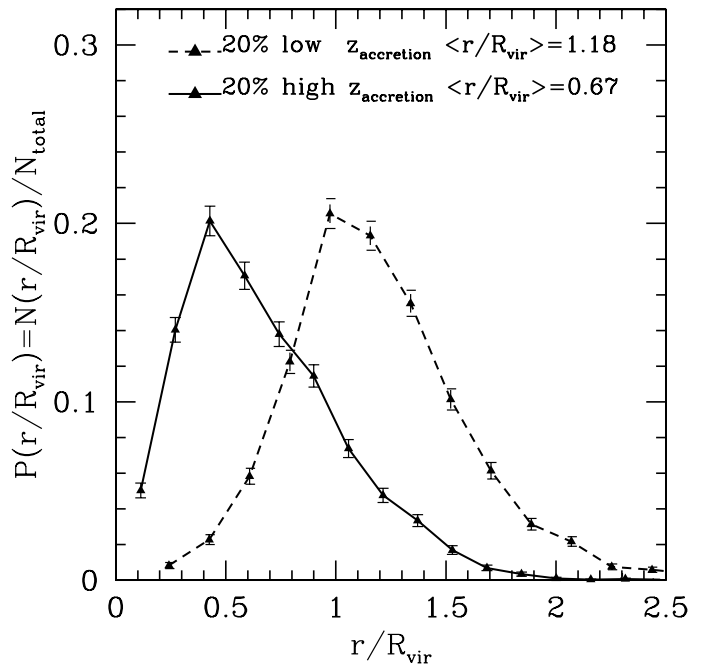,width=0.5\textwidth}}
  \caption{The radial distribution of sub-halos inside host halos at
    $z=0$. The solid line shows the 20\% of sub-halos with the
    earliest (high $z$) accretion times, while the dashed line shows
    the 20\% of sub-halos with the latest (low $z$) accretion times.}
  \label{radius_distri}
\end{figure}

To understand this puzzle, we have to note that our results are based
on subhalos only, while observations or the analysis of \cite{AB2010}
were based on galaxies.  To address the dependence of alignment on
accretion time of subhalos, we plot the radial distance of subhalos
with different accretion time in Figure~\ref{radius_distri}.  The solid
line shows that early accreted subhalos stay in the inner host halo,
but the late accreted subhalos stay near the virial radius of the
host. It is known that subhalos orbiting the inner host halo will
suffer from strong tidal forces, and thus are more likely to be
disrupted.  This effect is more efficient for subhalos with more
`radial' orbits, as they have more chance to pass the host center,
thus being more exposed to the strong tidal force and being disrupted
more efficiently \cite[e.g.,][]{Gan2010}. Thus only those with more
`circular' orbits will likely survive. Those subhalos with more
`circular' orbits will naturally produce a more isotropic
distribution.  For recently accreted subhalos, the effects of tidal
forces are still not strong enough to disrupt them as most of them are
still orbiting at the outer host halo.

Although the analysis of \cite{AB2010} made use of an $N$-body
simulation, they do include model galaxies given by the
semi-analytical model of \cite{Croton2006}. The `galaxies' are traced
by subhalos and unresolved subhalos (those disrupted ones). Thus the
main difference between \cite{AB2010} and ours is that they include
the so-called `orphan' galaxies \citep{Gao2004} which are not
associated with any resolved subhalos, but their positions can be
traced using the most-bound particles from the previous resolved
subhalos. These orphan galaxies can also be resolved in hydrodynamical
simulation with gas physics included as the condensation of baryons
boost the density distribution in subhalos, making it more resistant
to tidal disruptions \cite[e.g.,][]{Maccio2006}.

The `orphan' galaxies are mainly from early accreted subhalos, and
most of them are staying in the inner host halos, contributing
significantly to the total galaxy populations \citep{AB2010}. On one
hand, these orphan galaxies will include those subhalos with more
`radial' orbits, leading to a more aligned distribution. On the other
hand, orphan galaxies have orbited in the inner halo for longer time,
and they will follow the shape of halo more closely. Under the
assumption that central galaxies are also shaped dominantly by the
inner host halo, it will predict that red satellites align more strong
with central galaxies than the blue ones which are dominated by recent
accreted subhalos at the outer skirts of host halos. This will be
fully consistent with the observations.

\begin{figure*}
  \centerline{\psfig{figure=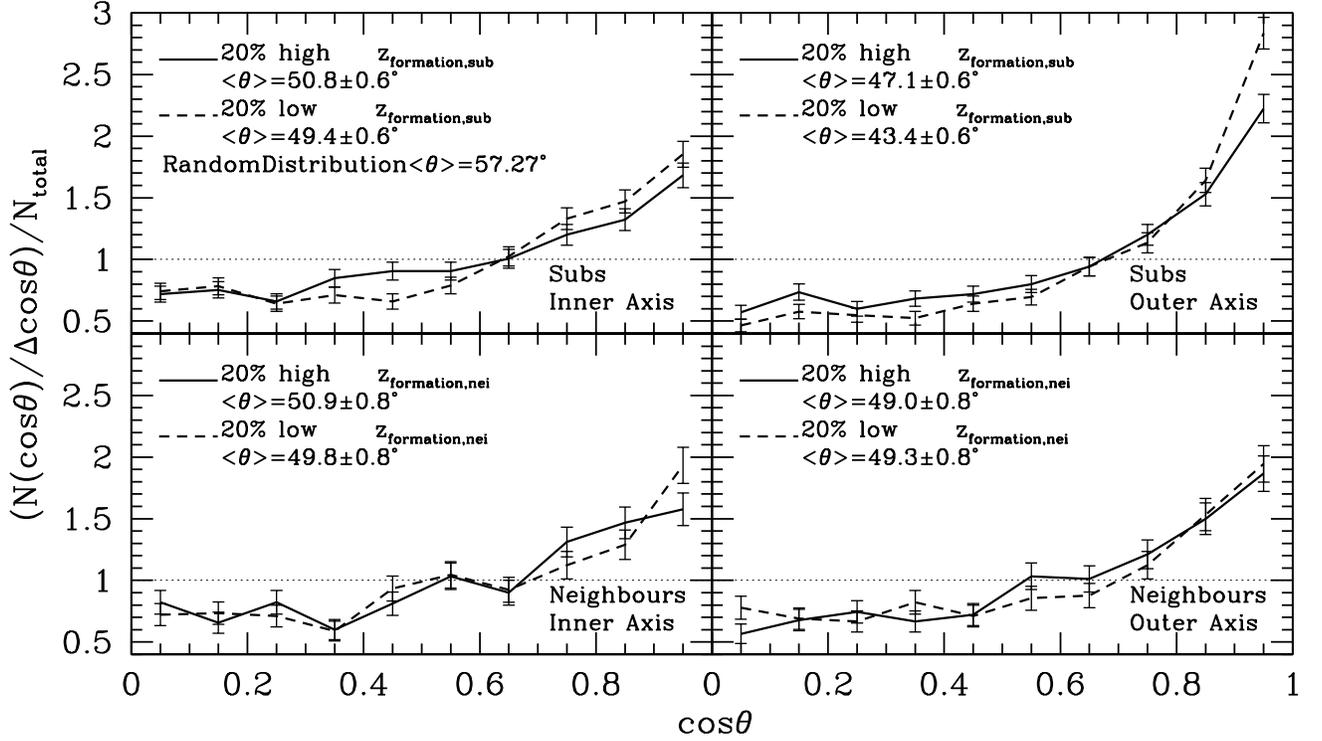,width=0.95\textwidth}}
  \caption{The dependence of angular distribution on formation
    redshifts for sub-halos (upper panels) and neighboring halos
    (bottom panels). The alignment along the ``\textit{inner axis}''
    and ``\textit{outer axis}'' are plotted in the left and right
    panels respectively. The solid line and dashed line are for
    sub-halos or neighboring halos with relatively high and low
    formation redshift respectively. The horizontal dotted line
    represents an isotropic angular distribution. \AD{Error bar
      indicates 1 $\sigma$ standard deviation in each bin.}  }
  \label{formation}
\end{figure*}

\begin{figure*}
  \centerline{\psfig{figure=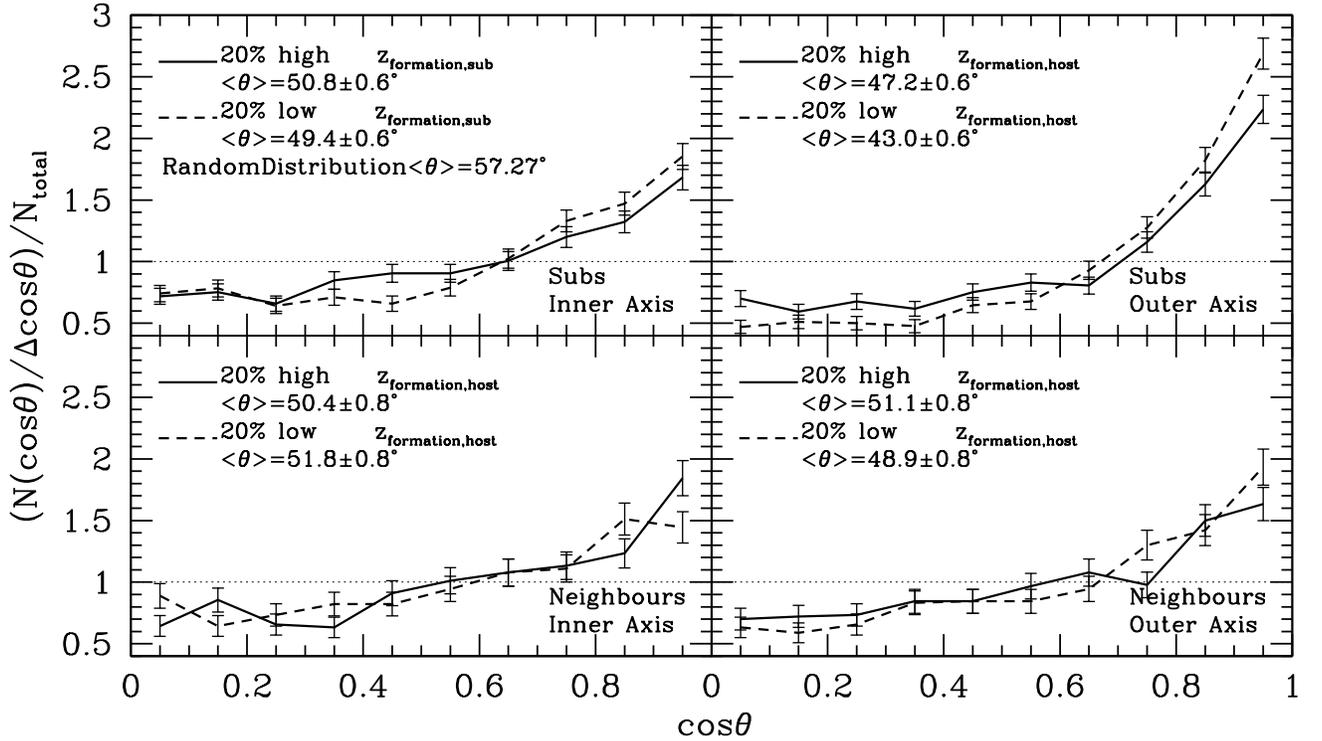,width=0.95\textwidth}}
  \caption{The influence of host halos' formation time on the angular
    distribution of sub-halos (upper panels) and neighboring halos
    (bottom panels). The other definition is the same as
    Figure~\ref{formation} except that the bins are divided in the
    formation time of host halos.}
  \label{formationhost}
\end{figure*}

Now we investigate the dependence of subhalos' alignment on their
formation time in Figure~\ref{formation}. Similar to the results shown in
Figure~\ref{accretion} the dependence of alignment of subhalos (upper
panels) is identical or slightly stronger in late formed
subhalos. This is due to the link between formation and accretion time
such that more early formed halos are more likely to be accreted at
earlier times. From the lower panel, it is found that for neighboring
halos, there is almost no dependence on their formation times.

Therefore, the influence of formation time is similar to that of
accretion time. In the bottom panels of Figure~\ref{formation}, the same
method is used as above to explore the influence of formation time on
neighboring halos. There is no obvious deviation between two lines in
these two panels. The formation time does not influence the
distribution of neighboring halos, different from the case for
sub-halos. The alignment signal should come from the coherent mass
distribution between halos and large scale structures, nothing to do
with the formation histories of neighboring halos.

Finally in this section, we show the effects of host halo formation
time on the alignment of subhalos/neighboring halos. It is seen from
Figure~\ref{formationhost} that the formation time of host halos have
weak effects on the alignment of subhalos/neighboring halos. From the
Press-Schechter(PS) theory \citep{PS1974}, and extended PS theory
(\citealt{LC1993, LC1994}; see also \citealt{Lin2003}), it is known
that the formation time of a halo depends on its mass such that more
massive halos form on average at later epochs. We should then expect
that the alignment of subhalos should be stronger in later-formed
halos, similar to the dependence seen in Figure~\ref{ang_host}. However,
it should be also noticed that for given halo mass, the formation time
has a broad distribution. Thus for a given formation time, there is a
wide distribution of halo masses which dilutes the dependence of
alignment on halo mass.


\section{CONCLUSION \& DISCUSSION}
\label{cha:concl}

Recent observations have found that satellite galaxies are not
randomly distributed, but rather they align with the major axis of the
central galaxy \cite[e.g.,][]{2005ApJ...628L.101B, Yang2006,
  2007MNRAS.376L..43A}. This intriguing result has generated great
interest, with many studies investigating the origin of this
phenomenon. The common conclusion from previous studies
\cite[e.g.,][]{AB2006, Kang2007} is that the alignment arises from the
non-spherical nature of dark matter halos in the cold dark matter
cosmology \cite[e.g.,][]{Jing2002}. However, there is no model which
can accurately predict the observed alignment signal and its
dependence on galaxy properties. The most difficult part of
theoretical modeling is in assigning the shape of central galaxies,
and most models do not agree on this aspect.

\begin{figure}
\centerline{\psfig{figure=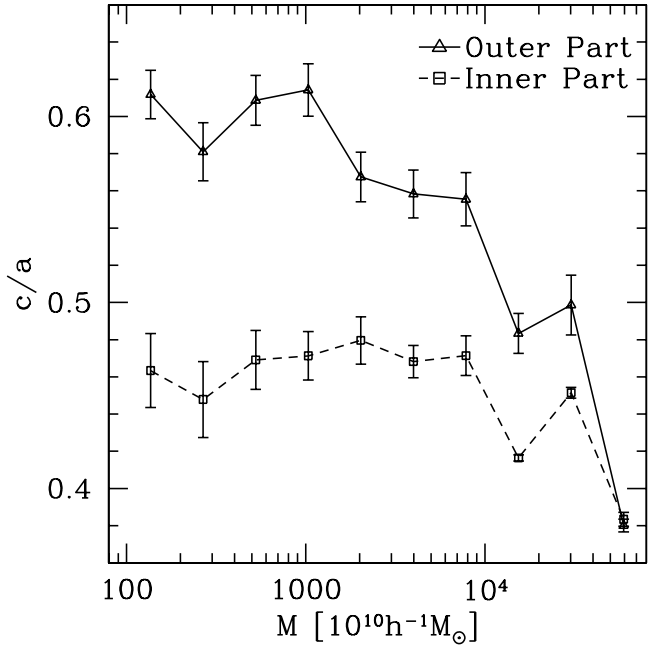,width=0.5\textwidth}}
\caption{Average axis ratio as a function of halo mass. Every point
  represent the average $c/a$ for a given mass bin. The error bars stand
  for 1 $\sigma$ standard deviation. All 2000 halos in our sample are
  included in this plot. }
    \label{ca}
    \end{figure}

In this paper, we re-visit the alignment problem using a cosmological
$N$-body simulation. Compared to previous studies, we focus on the
origin of the alignment with its dependence on the formation/accretion
of subhalos. We investigate if this dependence is from the assembly
bias or an evolution effect. Our results are summarized as follows,

\begin{itemize}

\item We use a new method to characterize the tri-axial halo shape
  following \cite{Jing2002}. Unlike the most widely used inertia
  tensor method which depends on the mass distribution within a given
  radius and is often contaminated by subhalos, the new method is able
  to determine the tri-axes of the halo at given local mass
  over-density, and excludes the effects on subhalos on the shape
  determination. We find that the measured halo shapes at different
  radii are well aligned. The mean alignment angle between the inner
  and outer part of halo is about $26.0^{\circ}, 38.9^{\circ}, 48.6
  ^{\circ} $ for host halos with $\Mvir \geq 10^{14} \Msun$, $10^{14}
  \Msun > \Mvir \geq 10^{13} \Msun$ and $ 10^{13} \Msun > \Mvir \geq
  10^{12} \Msun$ respectively.  The alignment between the inner and
  outer shapes is increasing with halo mass.

\item We study the alignment of both subhalos and neighboring halos
  around selected host halos.  Both subhalos and neighboring halos
  are found to align preferentially with the outer axes of host
  halos. Consistent with previous results \cite[e.g.,][]{Kang2007}, we
  find that the alignment of subhalos is stronger than the data if the
  outer axis of host halo is used for the shape of central
  galaxy. Better agreement with the data is achieved if the central
  galaxy follows the shape of host halo determined at the inner
  region. We also find that if the alignment between the central
  galaxy and the outer axis follows a Gaussian distribution with a
  mean of $0^{\circ}$ and a deviation of $25^{\circ}$, the predicted
  alignment also agree with the data.

\item The alignments of subhalos and neighboring halos depend on the
  mass of the host halos, such that more massive host halos have
  stronger alignment. This is due to the reasons that more massive
  halos are more flattened (embedded in filaments) and more massive
  halos have better alignment between its inner and outer axes. This
  is consistent with the observations that satellites in massive red
  central galaxies are more strongly aligned. \AD{In Fig~\ref{ca}, we
    show the dependence of halo flattening on the mass. It is seen
    that higher mass halos are more flattened (with lower c/a). Also
    found is that the inner region of halo is more flattened than the
    outer part. These results are consistent with previous studies
    \cite[e.g.,][]{Jing2002,2005ApJ...629..781K, Allgood2006, Maccio2008}. We also note
    that resolution (lower mass halos have fewer particles) has no
    effect on these results as the distribution shows consistent
    tendency up to the low-mass end in either case.  In Fig~\ref{ca},
    we include halos with their axes poorly detected. If these halos
    are excluded, the line of inner axes goes down at the low mass
    end.}  There is weak (if any) dependence of alignment on the mass
  of subhalos or neighboring halos themselves.

\item We study the alignment of subhalos with dependence on their
  formation time. It is found that there is no dependence along the
  inner major axes of host halos, and a strong dependence along the
  outer axes of the host halos such that the early accreted subhalos
  have lower alignment than the recent accreted ones. This is not
  consistent with the results of \cite{AB2010}, and is also
  inconsistent with observational evidence that red satellite galaxies
  (accreted more early) have stronger alignment with the central
  galaxy than blue satellites.

\end{itemize}

    The main contribution of this paper is that we find that the
    mis-alignment between the inner and outer axes of the dark matter
    halos can account for the observed alignment of satellite galaxies
    including the mass dependence. We confirm the results from most
    studies that the shape of the central galaxy cannot follow the
    shape of the whole dark matter halo, otherwise the predicted
    alignment signal is too strong. However, better agreement with
    observations can be obtained if the central galaxy follows the
    shape of the dark matter halo defined at the inner region
    (measured at an overdensity of $100\times 5^4 \rho_{\rm
      crit}$). In this case, the dependence of alignment on halo mass
    is also reproduced.

    Finally, we discuss whether the stronger alignment of red
    satellites due to the assembly bias or an evolution effect after
    accretion. As our simulation do not include models for galaxy
    formation, we use the subhalo population to address this
    question. Our conclusion is that the main contribution to the
    strong alignment of red satellites is from an evolution effect. If
    red satellites reside in subhalos that form at earlier times or
    having higher mass at accretion, Figure~\ref{ang_self} and
    Figure~\ref{formation} have shown that those subhalos do not have
    stronger alignment. Also the alignment of neighboring halos with
    higher mass or higher formation redshift are also identical. These
    results indicate that the stronger alignment of red satellites is
    not set at the time of accretion. On the contrary, their strong
    alignment should come from evolution effects after accretion.
    Actually, Figure~\ref{ang_comp} further shows that the alignment of
    neighboring halos is lower than the subhalo, implying that
    subhalos acquire stronger alignment after accretion.

    However, our results from Figure~\ref{accretion} seems not to support
    the evolution scenario. It shows that early accreted subhalos have
    lower alignment, indicating a negative evolution effect. It is
    also inconsistent with the results of \cite{AB2010} who have found
    that early accreted satellites have stronger alignment. We note
    that there are differences between our analysis and theirs. In our
    work, we do not have galaxies, but only subhalos. Also we do not
    include those disrupted subhalos (`orphan' galaxies in the model
    of \citealt{AB2010}). Thus in our work, those subhalos accreted at
    earlier times may have been disrupted, and this effect is more
    efficient for subhalos with more `radial' orbits as they will come
    to the host center and suffer strong tidal disruption. In addition,
    cautions should be taken here that the alignment on `inner' axes
    are determined less accurately because of limited resolution
    especially for low-mass halos and also for iso-density surfaces
    with small ellipticity (the latter one is also the case when
    determining the major axes observationally). Further investigation
    using simulations with higher resolution should be helpful.

    We simply conclude that to accurately predict the alignment of
    satellite galaxies found in observations and study its dependence
    on galaxy properties, we should use hydrodynamical simulation with
    gas physics and star formation, which can self-consistently
    predict the shape of central galaxies and the distribution of
    satellite galaxies around the central galaxy. Although current
    simulations with star formation still have difficulties to achieve
    better agreement with the data, we will show in an upcoming paper
    that the alignment of satellites and their color dependence can be
    better reproduced.

\acknowledgments

We thank the anonymous referee for useful suggestions to improve the
manuscript. And we thank Yipeng Jing, Xiaohu Yang, Cheng Li and Jiaxin
Han for helpful discussion and comments. WPL acknowledges supports by
the NSFC projects (No.10873027, 11121062, 11233005, U1331201).  XK is
supported by the NSFC (No. 11073055, 11333008), National basic
research program of China (2013CB834900) and the Bairen program of the
Chinese Academy of Sciences. The work is also supported by the
''Strategic Priority Research Program the Emergence of Cosmological
Structures'' of the Chinese Academy of Sciences，Grant No. XDB09010400.
The simulations were run in Shanghai
Supercomputer Center. Part of the computing were carried on the SHAO
Super Computing Platform.We acknowledge support from the MPG-CAS through
the partnership programme between the MPIA group lead by A. Macci\`o and
the PMO group lead by X. Kang

\end{document}